\documentclass[letterpaper]{article}

\usepackage{parskip}
\usepackage{graphicx}
\usepackage{amsmath}
\usepackage{caption}
\usepackage{chngpage}
\usepackage{amssymb}
\usepackage{textgreek}

\begin{document}
\title{Bank Density, Population Density, and Economic Deprivation Across the United States}
\author{Scott W. Hegerty\\
Department of Economics\\
Northeastern Illinois University\\
Chicago, IL 60625\\S-Hegerty@neiu.edu
}
\date{\today}
\maketitle

\begin{abstract}
Recent research on the geographic locations of bank branches in the United States has identified thresholds below which a given area can be considered to be a “banking desert.” Thus far, most analyses of the country as a whole have tended to focus on minimum distances from geographic areas to the nearest bank, while a recent density-based analysis focused only on the city of Chicago. As such, there is not yet a nationwide study of bank densities for the entire United States. This study calculates banks per square mile for U.S. Census tracts over ten different ranges of population density. One main finding is that bank density is sensitive to the measurement radius used (for example, density in urban areas can be calculated as the number of banks within two miles, while some rural areas require a 20-mile radius). This study then compiles a set of lower 5- and 10-percent thresholds that might be used to identify “banking deserts” in various urban, suburban, and rural areas; these largely conform to the findings of previous analyses. Finally, adjusting for population density using regression residuals, this paper examines whether an index of economic deprivation is significantly higher in the five percent of “desert” tracts than in the remaining 95 percent. The differences are largest—and highly significant—in the densest tracts in large urban areas.  
\end{abstract}

\section{Introduction}
Although the use of online and mobile banking has increased markedly in recent years, a case can be made that “bank branches matter,” particularly for older or lower-income residents with limited mobility or technological fluency. A recent report by the Federal Reserve’s Board of Goverrnors (2019) notes that more than half of rural counties lost banks between 2012 and 2017, and that residents face difficulties adjusting to these changes. Likewise, the Woodstock Institute (2018) notes the challenges faced by older residents in the Chicago area.	 

Geographic areas with relatively little financial access— so-called “banking deserts” —might instead be served by alternative (and sometimes predatory) financial service providers. In addition, they have been linked to social effects such as crime (Kubrin \textit{et al.}, 2011) or adverse public health outcomes (Eisenberg-Guyot \textit{et al.}, 2018). Economically, residents of areas with few banks often pay higher interest rates for limited credit (Ergungor, 2010; Nguyen, 2019). This is usually explained by a lack of information among lenders who do not understand the communities in which they are lending, or to high monitoring costs due to increased distance between borrowers and lenders (Degryse and Ongena, 2005). 

Much of the literature that finds gaps in bank service provision—the so-called “spatial void” hypothesis noted by Smith \textit{et al.} (2008)—focuses on the payday loan operators and other “fringe banking” providers that seek to exploit these gaps and offer limited services at much higher costs to the consumer. Brennan \textit{et al.} (2011), for example, find that poor neighborhoods in Winnipeg have been overserved by these non-bank financial institutions and underserved by banks and credit unions. Chen  \textit{et al.} (2014), on the other hand, find no evidence of any “spatial void” in their study of U.S counties. Racial disparities are an important covariate with the lack of banking services (Wheatley, 2010; Cover \textit{et al.}, 2011) as are income and socioeconomic variables such as housing tenure (Hegerty, 2016;  Dunham \textit{et al.}, 2018). 

Using a distance-based measure of “banking deserts,” Kashian \textit{et al.} (2018) examine more than 60,000 Census tracts from 2009 to 2015, controlling for population density by regressing the distance from each centroid to its nearest bank and examining the residuals for urban, rural, and suburban areas separately. They find that poverty is negatively related to bank proximity only in urban areas; this relationship is insignificant in suburban tracts and positive in rural ones.

Most recently, Hegerty (2019b) conducts an analysis of bank locations in Chicago, focusing on bank counts within one and two miles of each block-group centroid. He finds “banking deserts” in roughly nine percent of the city, and estimates that these areas contain roughly 0.4 banks per square mile. These block groups are shown to both be poorer and to have fewer white residents than the city as a whole, and particularly in comparison to neighborhoods with large shares of bank branches. But the “rule of thumb” implied in the paper most likely only applies to large cities, so further empirical work could find bank densities for other locations.

Is it possible to refine this definition to create a nationally applicable threshold (or set of thresholds), and how might these cutoffs differ in large cities, smaller cities, and rural areas? Kashian \textit{et al.} (2018) consider the lowest 5 percent of residuals from a regression on population density (their measure of “adjusted” bank density) and find cutoffs of 1.56 miles for urban areas, 4.28 miles for suburban areas, and 12.54 miles for rural areas. Assuming that this implies one bank on the edge of each circle with the given radii, this converts to 0.41 banks per square miles only for urban areas; the corresponding suburban and rural values are 0.05 and 0.006, respectively. If banks were uniformly distributed in rural areas, this latter value would imply only two banks in a 10-mile radius, or less than eight within a 20-mile radius. 

Noting that the relationship between bank and population densities appears to be nonlinear, this study controls for population density by performing separate regressions over 10 different density ranges. Adjusted bank densities are highly correlated with the unadjusted densities, so the latter are the primary focus here.  Comparing these results with those of Hegerty (2019b), the 80\textsuperscript{th} and 90\textsuperscript{th} percentiles—in which most Chicago tracts are located—are found to have  bottom 5 and 10 percent thresholds that match both the earlier block-group-level analysis and a tract-level reestimation conducted here. Many of the most rural areas have no banks even within 20 miles, but many rural and suburban areas have density thresholds in line with the results of Kashian \textit{et al.} (2018). Comparing bank densities with the tract-level economic deprivation index of Hegerty (2019a), deprivation is shown to have significantly higher in the 5 percent of tracts with the lowest bank densities than in the remaining 95 percent, but that these differences are largest at higher population densities.   

\section{Methodology}
Bank data are taken from the FDIC’s Summary of Deposits database; they were current as of June 30, 2019. These provided data contained 87,931 data points with both XY coordinate data and address data in the lower 48 states and the District of Columbia. These data points, which were able to be plotted in Geographic Information Systems software, comprise more than 99.9 percent of the original dataset. To maintain a single method of geocoding, no attempt was made to add the additional bank locations to the database. In addition, while Hegerty (2019b) notes that FDIC latitude and longitude data contain various inaccuracies—and some were found in the current study—these were taken as is.

These are plotted against the 71,593 census tracts in 48 states plus DC. These tracts are used for two reasons. First, their centroids serve as the basis for buffers of different radii, within which bank densities are calculated. Second, these densities are compared against an index of economic deprivation proposed by Hegerty (2019a), which uses U.S. Census data (2015 ACS 5-year estimates) to combine five socioeconomic variables into a single measure.  

\begin{figure}[ht]
\hfill
\caption{Bank Locations (2018) and Population Density for U.S. Census Tracts.}
\includegraphics[width=1\textwidth]{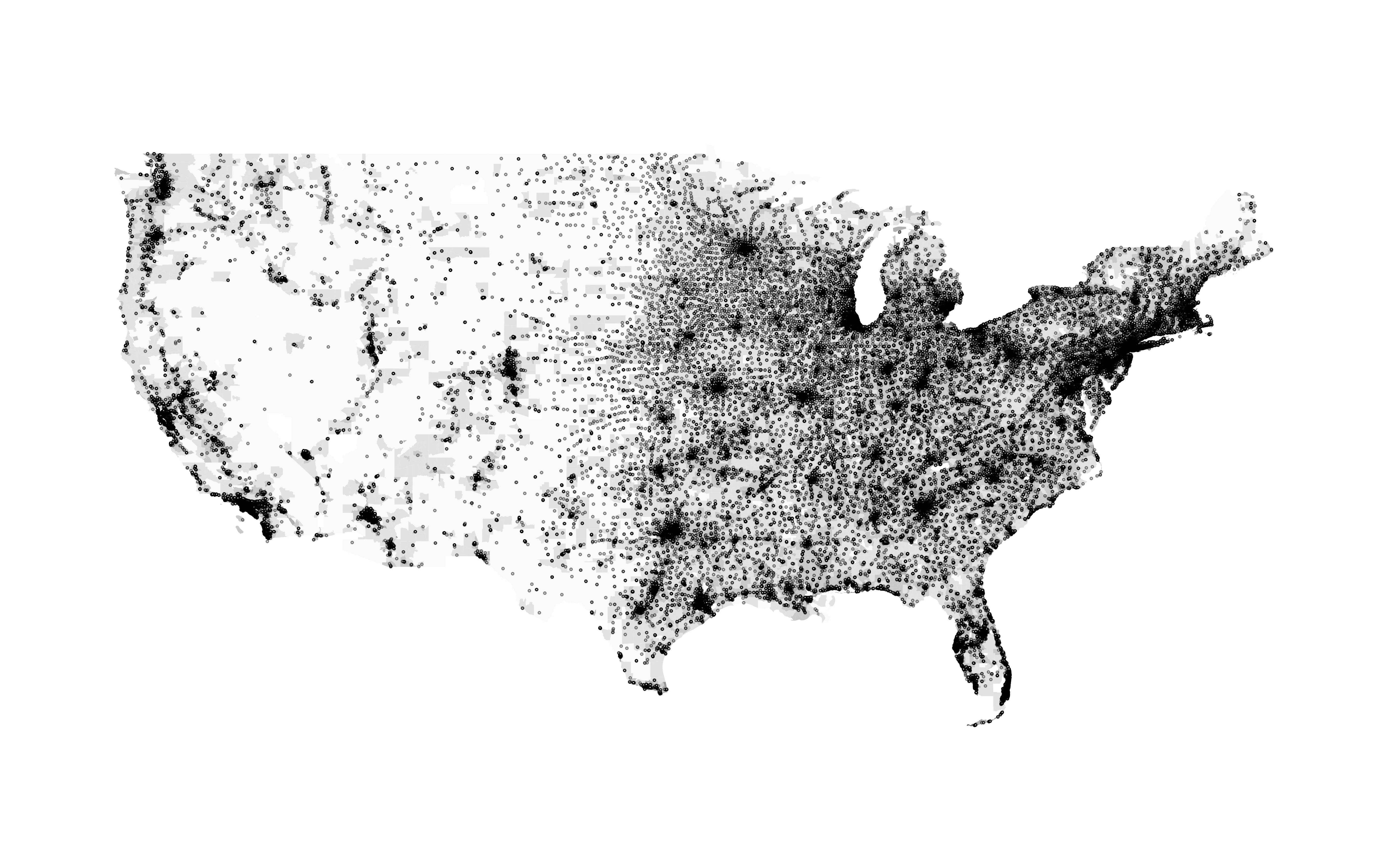}
\caption*{Choropleth map of population density with natural breaks:\\Light = low density; black = higher.\\
Census data: 2015 American Community Survey 5-year estimates.}
\end{figure}

In this study, bank densities are calculated as the number of bank locations per square mile, within the given radii from each tract centroid. For example, a tract with 20 banks in a 2-mile radius (which has an area of 12.57 square miles) would have 1.59 banks per 1 square mile. If, as the radius increases, the number of nearby banks changes at a rate different from the circle area, this density could increase or decrease. It is therefore possible to look for some sort of “optimal” radius for future analyses, as well as to use different measures for rural and urban areas. While densities are calculated at quarter-mile increments from 0.25 to 20.00 miles, the primary measures of analysis are the 2-, 5-, 10, and 20-mile radii.

To control for different types of urban, suburban, and rural areas, this paper examines and controls for population density. After plotting bank density versus (log) population density and finding a distinct nonlinear relationship, a procedure similar to that of Kashian \textit{et al.} (2018) is applied, which uses regression residuals as a measure of adjusted bank density. Log population densities are split into 10 equal segments (with unequal numbers of tracts, however) and separate OLS regressions are performed for each. The resulting residuals serve as a measure of “adjusted” bank density that can be used to measure relationships with a number of socioeconomic variables. For both the adjusted and the unadjusted bank densities, this paper focuses on the 5 and 10 percent thresholds, within each decile, using the four major measurement radii.

Finally, economic deprivation is compared in these low-density “banking deserts” against deprivation in the remaining tracts. Mean deprivation in the bottom 5 percent and top 95 percent of tracts within each population density decile are calculated and tested for significant differences using standard t-tests. Overall, the differences are largest and most significant in more “urban” tracts, generally at the 70\textsuperscript{th} percentile of log population density or higher.

\begin{figure}[ht]
\hfill
\caption{Bank Density Quantiles at Different Measurement Radii (in Miles).}
\includegraphics[width=1\textwidth]{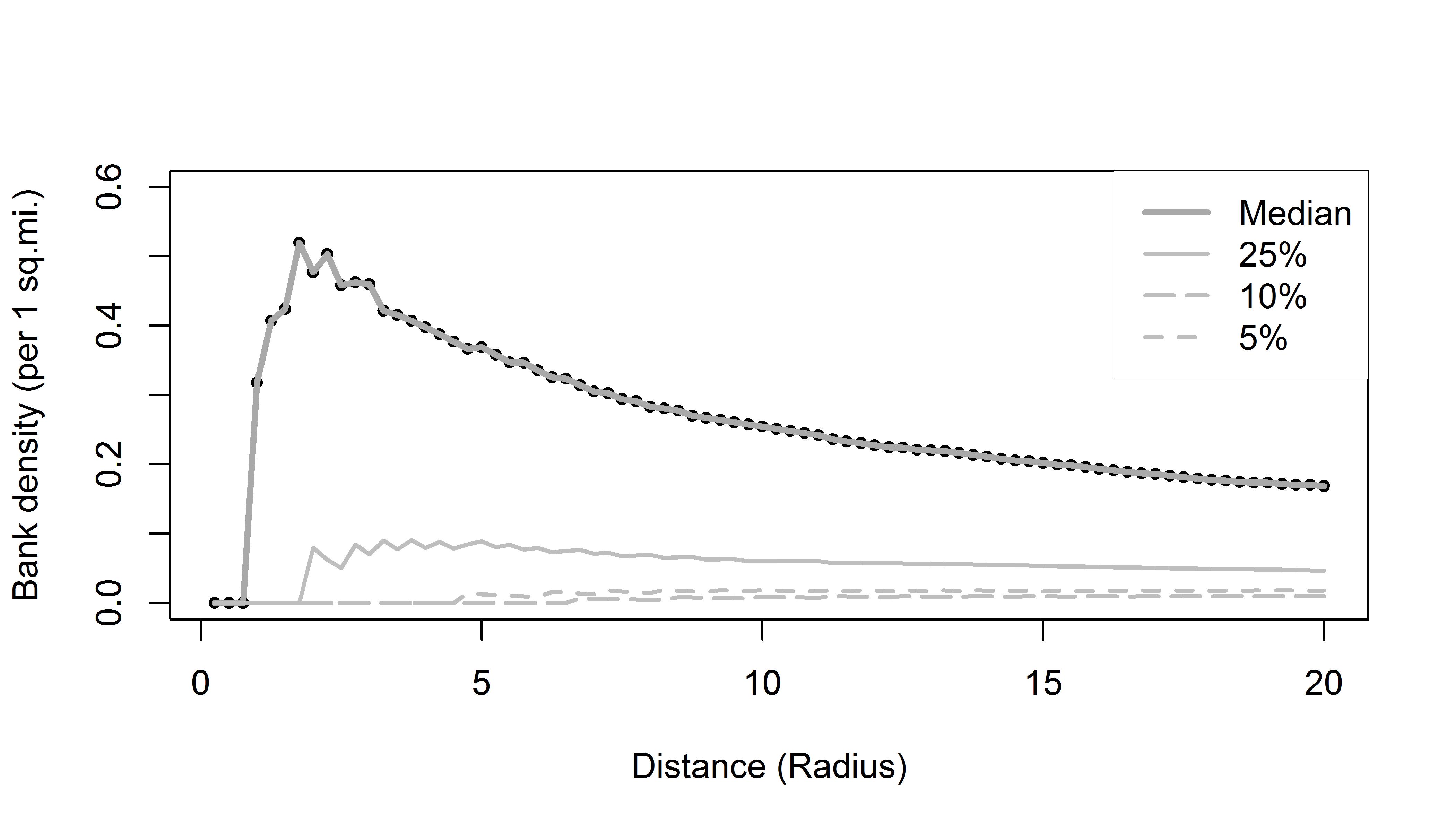}
\end{figure}

\section{Results}
Figure 1 shows that, as expected, banks are clustered in urban areas. In the sparsely populated Western states, there are relatively few banks, even in tracts with higher population densities. Most likely, many areas will have zero banks per square mile, even within a large radius, but tracts with higher population densities and no nearby banks are more likely to be considered to be “banking deserts,” since these values are lower than expected.

\begin{table}[ht]
\caption{Summary Statistics for Bank Densities and Deprivation Index.}
        \begin{center}
\begin{tabular}{lrrrrrr}

&1-Mile&2-Mile&5-Mile&10-Mile&20-Mile&Deprivation\\
\hline
Min.&0.00&0.00&0.00&0.00&0.00&-2.43\\
1Q&0.00&0.08&0.09&0.06&0.05&-1.26\\
Median&0.32&0.48&0.37&0.25&0.17&-0.45\\
Mean&1.10&0.94&0.72&0.54&0.35&-0.04\\
3Q&1.27&1.11&0.83&0.64&0.42&0.77\\
Max.&84.99&34.85&12.22&5.81&2.62&11.55\\
Area&3.14&12.57&78.54&314.16&1256.64&\\
\hline
\end{tabular}
\end{center}
\caption*{Values calculated per 1 square mile.\\
Areas of circles with given radii are presented in square miles.}
\end{table}

Figure 2 shows how the choice of measurement radius affects the calculation of bank density. The highest median bank density is calculated when a radius of around 2 miles is applied—the average from 1.75 to 2.25 miles is almost exactly 0.500. This density declines as the measurement radius increases; most likely, the resulting circles exceed the size of nearby urbanized areas, so as the radius increases, the rate of new banks falling within the buffer is less than the increase in total area. The 5 and 10 percent quantiles stabilize for radii larger than 5 miles, however. Table 1 presents summary statistics for this sample of (unadjusted) banking densities at the major radii used for this study, as well as for the index of economic deprivation. Since the median exceeds the mean in all cases, the data are skewed right.

\begin{figure}[ht]
\hfill
\caption{Bank Density Versus Population Density (2-mile Measurement Radius).}
\includegraphics[width=1\textwidth]{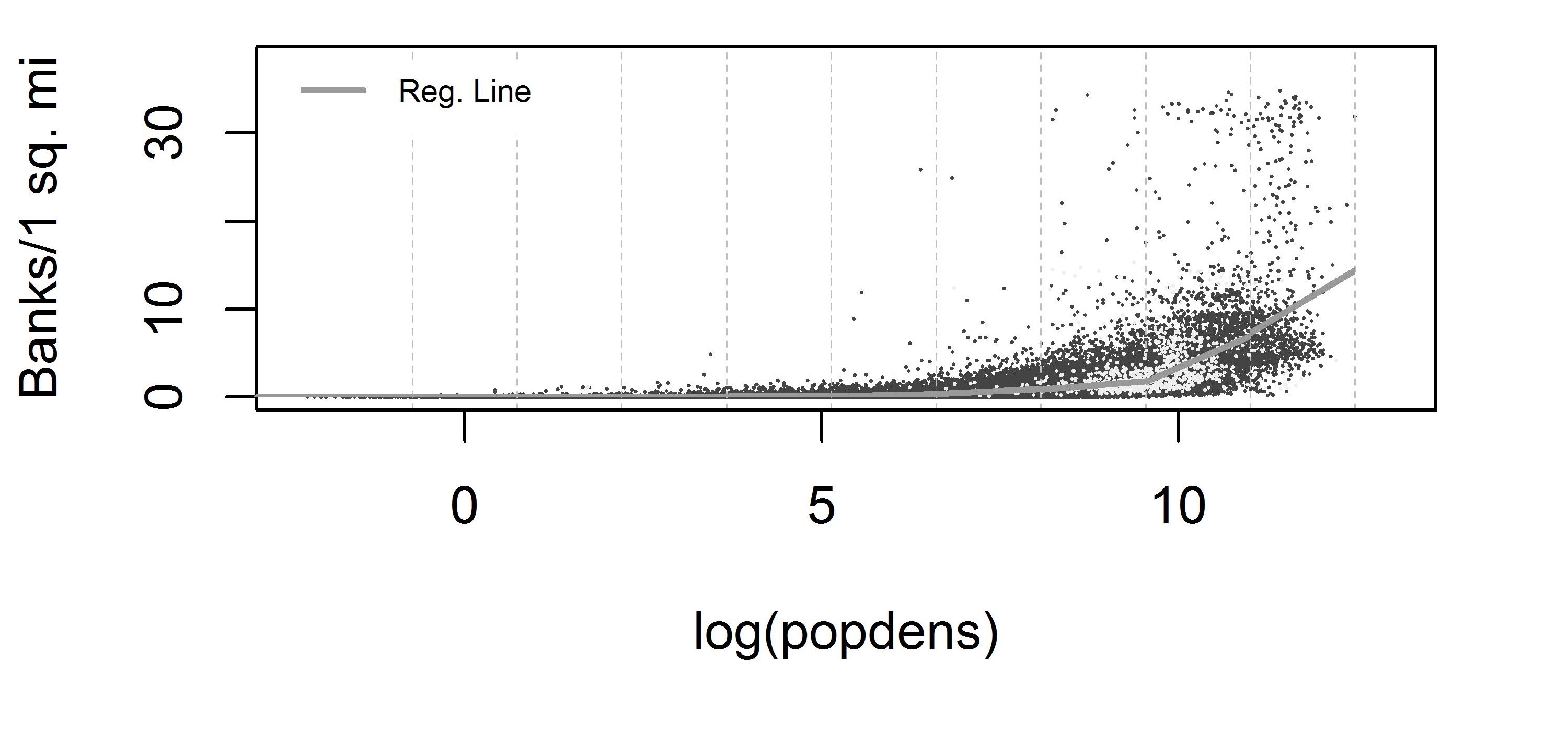}
\caption*{Vertical lines: Separators between density ranges; OLS regressions conducted for 10 separate groups.\\
Grey line: OLS regression lines within each segment.\\
Grey dots: Tracts located in the city of Chicago.
}
\end{figure}

Figure 3 plots bank densities (at 2 miles) against log population density; this relationship is nonlinear even when various additional transformations (not shown here) are applied. The highest decile (100\textsuperscript{th} percentile) of log population density is typically composed of tracts in New York City, and the lowest (10\textsuperscript{th} percentile) contains tracts in states such as Idaho and Wyoming. For comparison purposes, Chicago tracts are depicted in grey; these are typically located in the 80\textsuperscript{th} and 90\textsuperscript{th} percentile ranges.  Table 2 presents statistics on the relative size, population, area, and population density of each decile. The 70\textsuperscript{th} and 80\textsuperscript{th} percentiles (7th and 8th deciles) contain the largest shares of tracts and population in a very small land area. The 40\textsuperscript{th} to 60\textsuperscript{th} percentiles are relatively less dense and might represent more small-town or “suburban” areas. Since these segments are density-, not geographically-based, they might also represent parts of a major city. This study, however, does not focus on specific locations. In fact, as is shown for Chicago, one city’s component tracts might fall across multiple density segments.

\begin{table}[ht]
\caption{Population, Area, and Population Density by Tract Group.}
\begin{tabular}{rrrrrrrrr}

&&\multicolumn{3}{l}{Cumulative \%}&\multicolumn{4}{l}{log(popdens)}\\
Decile&Size&Size&Area&Pop&Min&Max&exp(Min)&exp(Max)\\
\hline
All&71584&&&&&&&\\
\hline
10&68&0.1&6.77&0.02&-10.00&-0.74&0.0&0.5\\
20&368&0.6&26.48&0.25&-0.74&0.73&0.5&2.1\\
30&1396&2.6&51.89&1.44&0.73&2.20&2.1&9.0\\
40&4867&9.4&78.19&6.72&2.20&3.67&9.0&39.2\\
50&8234&20.9&93.43&18.18&3.67&5.14&39.2&170.0\\
60&9310&33.9&97.60&32.00&5.14&6.60&170.0&738.0\\
70&19243&60.7&99.37&59.51&6.60&8.07&738.0&3203.5\\
80&22561&92.3&99.97&91.86&8.07&9.54&3203.5&13904.9\\
90&4772&98.9&100.00&98.59&9.54&11.01&13904.9&60355.1\\
100&765&100&100&99.99&11.01&12.48&60355.1&261973.9\\
\hline

\end{tabular}

\end{table}

\begin{table}[ht]
\caption{Bank Density at Lower Quantiles Within Density Deciles and at Different Radii.}
  \begin{adjustwidth}{-.5in}{-.5in}  
        \begin{center}
\begin{tabular}{rrrrrrrrrrrrr}

&\multicolumn{3}{l}{2-mile}&\multicolumn{3}{l}{5-mile}&\multicolumn{3}{l}{10-mile}&\multicolumn{3}{l}{20-mile}\\

Decile&\%Zero&5\%&10\%&\%Zero&5\%&10\%&\%Zero&5\%&10\%&\%Zero&5\%&10\%\\
\hline
All&24.3&0&0&8.8&0&0.013&1.8&0.010&0.019&0.4&0.010&0.018\\
\hline
10&95.6&0&0&79.4&0&0&66.2&0&0&45.6&0&0\\
20&92.9&0&0&79.6&0&0&61.1&0&0&27.2&0&0\\
30&90.3&0&0&69.9&0&0&35.6&0&0&7.2&0&0.001\\
40&88.0&0&0&53.3&0&0&8.6&0&0.003&0.4&0.004&0.006\\
50&76.6&0&0&25.6&0&0&1.0&0.006&0.010&0&0.010&0.014\\
60&41.2&0&0&2.7&0.013&0.025&0.1&0.016&0.022&0&0.011&0.018\\
70&5.9&0&0.080&0.1&0.076&0.115&0&0.035&0.057&0&0.021&0.034\\
80&0.7&0.239&0.318&0&0.191&0.280&0&0.095&0.150&0&0.047&0.072\\
90&0&0.557&0.796&0&0.573&0.789&0&0.400&0.605&0&0.220&0.347\\
100&0&2.467&3.501&0&2.422&3.002&0&1.360&2.019&0&0.573&1.028\\
\hline

\end{tabular}
        \end{center}
    \end{adjustwidth}

\caption*{Unadjusted values = banks per 1 square mile within the given radius from tract centroids.\\
\%Zero = percentage of tracts with no banks within a given radius.}

\end{table}

Table 3 shows the distributions of bank densities within each segment, as well as for the entire sample. Many tracts have zero banks per square mile; while this is especially true for the 2- and 5-mile radii, it also applies for the least-dense deciles when a 20-mile radius is applied. A two-mile measurement radius, therefore, might be useful for a city such as Chicago, but it fails to find even a single bank in tracts at the 60\textsuperscript{th} percentile for log population density and below. The same can be said for the 5-mile radius (with a large share of zero-density tracts at the 50\textsuperscript{th} percentile and below). The 10-and 20-mile radius see the same effect for the 30\textsuperscript{th} and 20\textsuperscript{th} percentiles, respectively. 

\begin{table}[ht]
\caption{Lower Quantile Bank Densities for Tracts Located in Chicago (N = 781).}
        \begin{center}
\begin{tabular}{rrrr}

Radius (mi.)&1\%&5\%&10\%\\
\hline
1&0&0.318&0.637\\
2&0.382&0.716&0.875\\
5&0.545&0.764&0.955\\
10&0.632&0.907&1.044\\
20&0.813&0.903&0.963\\

\hline
\end{tabular}
        \end{center}
\end{table}

Clearly the choice of distance range matters among urban, suburban, and rural areas when calculating bank densities. Population density also affects the thresholds that can be used when defining “banking deserts.” In Table 3, the 5 and 10 percent quantile values are shown to differ across population density deciles. Chicago’s range (the 80\textsuperscript{th} and 90\textsuperscript{th} percentiles) match the findings of Hegerty (2019b), who indirectly calculated a density of roughly 0.4 banks per square mile, using 1- and 2-mile radii, for the bottom nine percent of tracts. These findings are also confirmed for the 781 tracts in the city of Chicago, which are presented in Table 4, particularly when a 1-mile radius is applied. For more suburban ranges, using a 20-mile measurement radius, these quantiles fall in a range from 0.005 to 0.015 banks per square mile (which translates to between roughly 6 and 20 banks within this very large circle), and aligns with the results of Kashian \textit{et al.} (2018).

\begin{table}[ht]
\caption{Summary for “Banking Desert” Thresholds for Selected Population Densities.}
        \begin{center}
\begin{tabular}{lrrr}

Type&Percentile&Banks/mi2&\#Banks in Radius (Radius)\\
\hline
Urban&80-90&0.24-0.80&3-10 (2mi)\\
Less Urban&60-70&0.013-0.115&2-9 (5mi)\\
Rural&40-50&0.004-0.14&5-17 (20mi)\\
Very Rural&$\leq$30&0&0 (20mi)\\
\hline

\end{tabular}
        \end{center}
\caption*{Derived from 5 and 10 percent quantiles in Table 3.}
\end{table}

Table 5 summarizes the thresholds for the lowest bank densities in each of four density categories. The lowest threshold in urban areas is double the threshold for “less urban” areas with population densities below 3,200 per square mile. The lowest 5 percent of rural tracts have even lower bank densities. Converted into the number of banks within a given radius, these figures show (for example) that to find nine banks, one might need a circle with a 2-mile radius in an “urban” area; five miles in a less-urban area, and 20 miles in a rural area. These numbers can serve as a basis when assessing whether specific neighborhoods or areas can be classified as such a “desert.” 

The relationship between population density and bank density becomes more pronounced as the former variable increases. This is shown via the slope coefficients from a bivariate OLS regressions, within each decile. Depicted visually in Figure 3 for the 2-mile radius, they are generally significant at the 40\textsuperscript{th} percentile or greater regardless of the measurement radius. For larger cities, we expect that even the harshest “desert” is expected to have at least a few banks nearby.

\begin{table}[ht]
\caption{Statistics for Regression Errors and “Bank Deserts” at Different Radii.}
        \begin{center}

\begin{tabular}{rrrrrrrrr}

&\multicolumn{2}{l}{2-mile}&\multicolumn{2}{l}{5-mile}&\multicolumn{2}{l}{10-mile}&\multicolumn{2}{l}{20-mile}\\
Decile&\textrho&T-Test&\textrho&T-Test&\textrho&T-Test&\textrho&T-Test\\
\hline
All&0.610&\textbf{16.742}&0.917&\textbf{24.449}&0.984&\textbf{32.475}&0.908&\textbf{34.068}\\
10&0.356&-0.467&0.706&-0.977&0.673&-0.977&0.771&-0.977\\
20&0.444&-0.118&0.619&-1.818&0.609&-0.938&0.421&1.208\\
30&0.514&1.840&0.809&\textbf{2.923}&0.917&\textbf{4.252}&0.824&\textbf{5.035}\\
40&0.563&0.630&0.896&\textbf{2.263}&0.824&\textbf{5.794}&0.801&\textbf{10.311}\\
50&0.741&-2.637&0.808&\textbf{2.369}&0.814&\textbf{10.797}&0.867&\textbf{10.665}\\
60&0.853&-2.311&0.829&\textbf{8.596}&0.904&\textbf{13.747}&0.941&\textbf{12.880}\\
70&0.915&\textbf{8.089}&0.923&\textbf{13.846}&0.941&1\textbf{7.482}&0.956&\textbf{16.852}\\
80&0.919&\textbf{19.288}&0.875&\textbf{19.102}&0.880&\textbf{23.005}&0.885&\textbf{25.499}\\
90&0.764&\textbf{10.775}&0.747&\textbf{9.428}&0.820&\textbf{6.607}&0.889&\textbf{6.130}\\
100&0.917&\textbf{3.546}&0.984&\textbf{3.316}&0.908&\textbf{2.457}&0.610&\textbf{2.884}\\
\hline

\end{tabular}
        \end{center}

\caption*{
\textrho  = Spearman rank correlations between adjusted and unadjusted measures.\\
T-Test for significant differences in mean deprivation scores between the 5\% "desert" tracts and 95\% remaining tracts within each decile.\\
\textbf{Bold} = significantly positive at 5 percent.
}

\end{table}

Table 6 shows large correlations between the adjusted and unadjusted bank-density measures, with the highest Spearman coefficients found around the 70\textsuperscript{th} population-density percentile. Because they will have very few zero values at any radius, while at the same time differentiating between an area with no bank access within a populated area and a similar area in a sparsely populated one, the adjusted bank density measures are preferred when examining associations with other socioeconomic variables.

\begin{figure}[ht]
\hfill
\caption{Mean Deprivation Values, “Bank Desert” vs. non-Desert Tracts.}

\includegraphics[width=.5\textwidth]{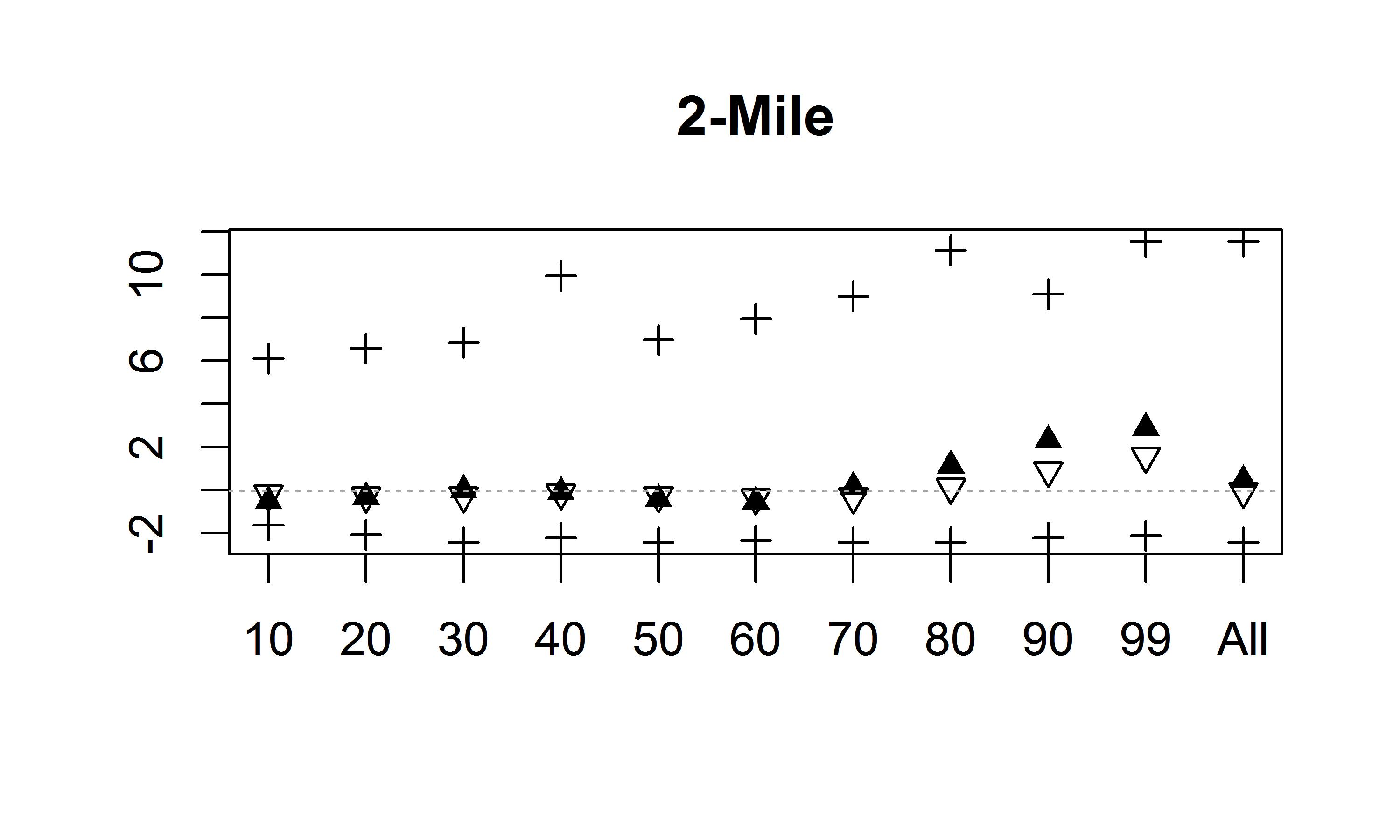}
\includegraphics[width=.5\textwidth]{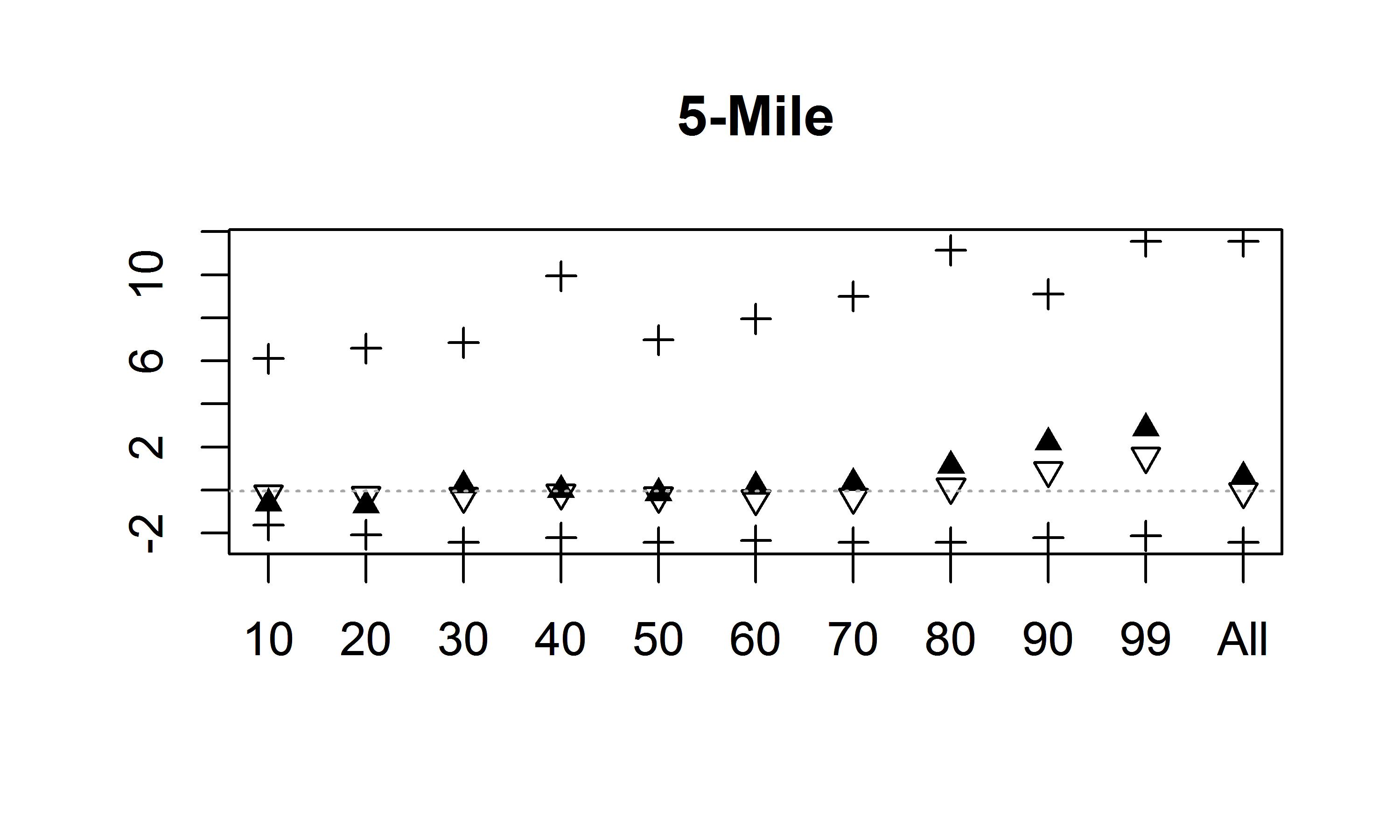}
\includegraphics[width=.5\textwidth]{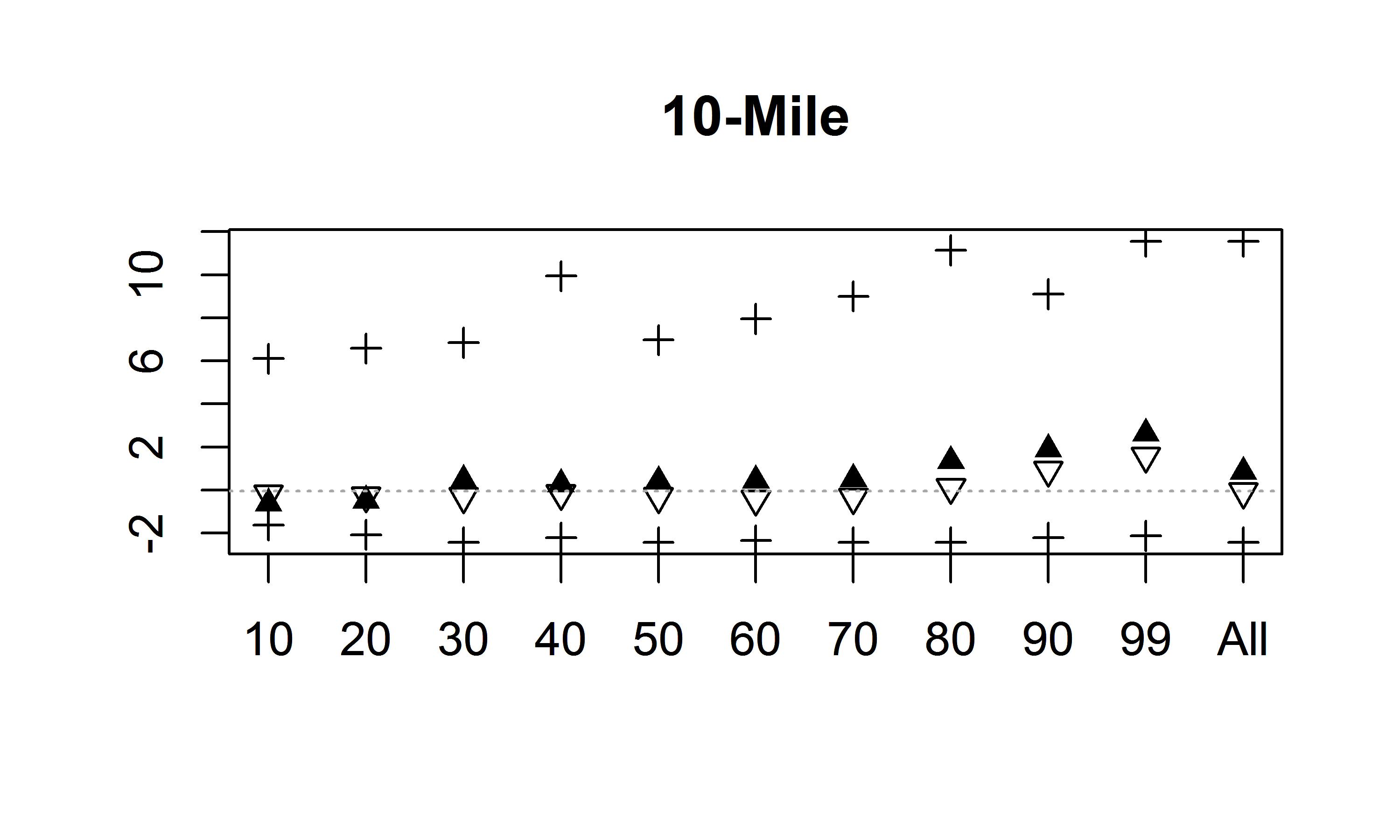}
\includegraphics[width=.5\textwidth]{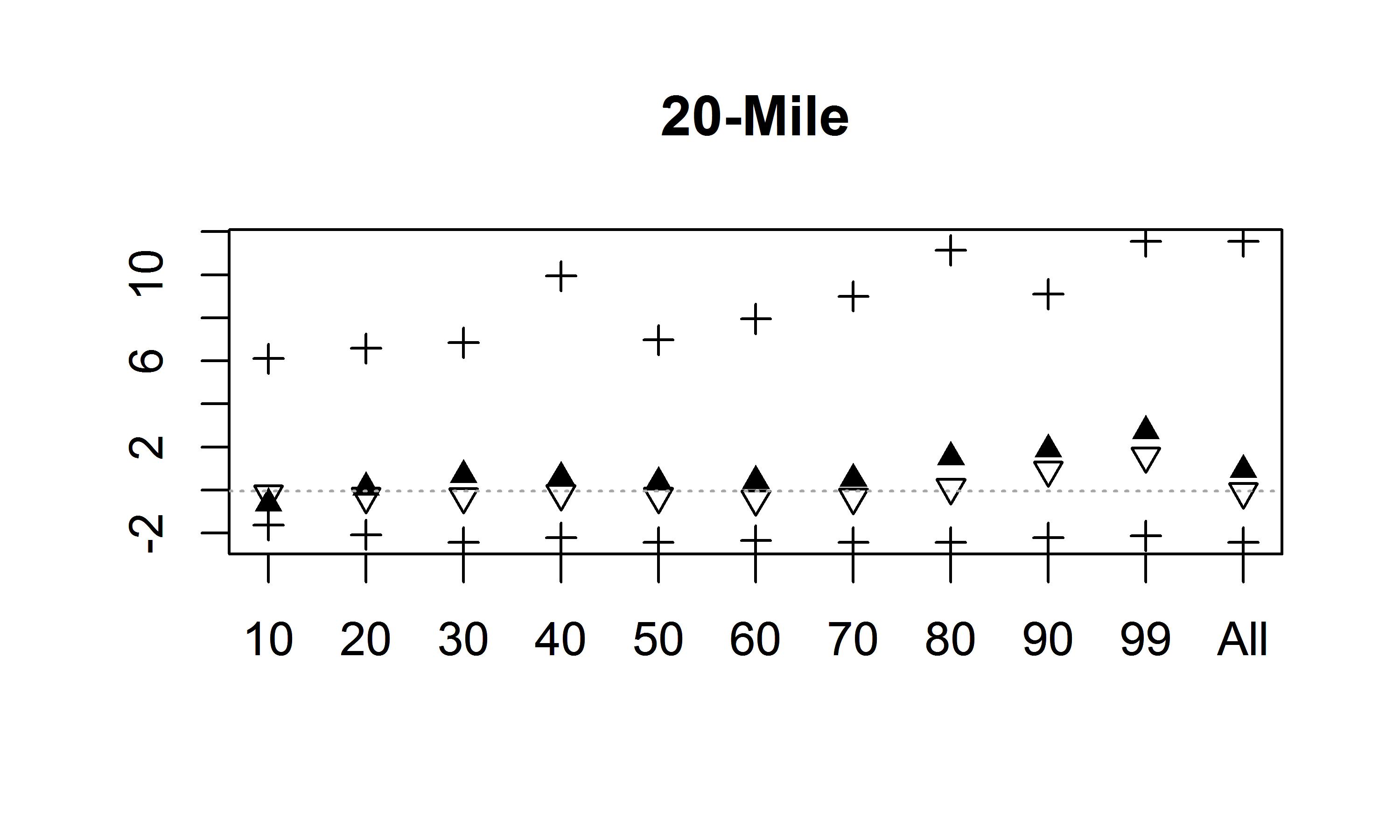}
\caption*{
$ \blacktriangle$ = Bank Deserts (lower 5\% of adjusted bank densities)\\
$\triangledown$ = Remaining 95\% of tracts	\\ 				
+ = Minimum, Maximum deprivation values	\\
Horizontal line = mean deprivation value
}

\end{figure}

These adjusted values are used to compare differences in socioeconomic deprivation between “banking deserts” and non-deserts. Depicted graphically in Figure 4, the gaps between the bottom 5 percent of tracts and the remaining 95 percent are largest at the 80\textsuperscript{th} percentile and above. The image of a socioeconomically deprived neighborhood with limited banking access, therefore, might hold mainly for large cities as Chicago. The t-tests in Table 6 show that, while small, these differences in mean deprivation scores for most tracts and density deciles are indeed significant. While differences are only significant for the 70\textsuperscript{th} percentile and above within a 2-mile radius, they are significant for the 30\textsuperscript{th} percentile and above when larger radii are used. Overall, areas with relatively few banks have significantly higher levels of economic deprivation than do non-deserts, but both groups’ levels—as well as the differences between them—are only large in highly urban areas.

\section{Conclusion}
While “banking deserts” are often examined by researchers and advocates who wish to increase communities’ access to financial services and limit the reach of high-interest alternative providers, little has been done to provide an exact definition of such an area. Combining the approaches of a recent distance-based analysis and a city-level density-based approach, this study calculates bank densities (number of branches per square mile) for Census tracts across the lower 48 states and the District of Columbia. Because banks’ service areas often exceed tract boundaries, these densities are calculated using a variety of measurement radii for each tract and centroid. Calculated densities, however, are sensitive to the choice of radii—a choice of two miles gives the highest median density nationwide and works well for large cities, while larger radii of 5, 10, or 20 miles work well in less-dense areas.

	Bank density is compared against population density, but because the relationship is nonlinear, comparisons are made within ten deciles based on log population density. Linear regression suggests that this relationship is strongest for the highest density, most urban, tracts. Regression analysis is also used to create an “adjusted” bank density measure, which controls for population density within each decile. 

	In line with the work of Kashian \textit{et al.} (2018) and Hegerty (2019b), this study defines thresholds to define “banking deserts” and which are based on the lowest 5 and 10 percent of unadjusted banking densities within each decile. These correspond to roughly 0.24 to 0.80 banks per square mile in highly urban tracts, 0.013 to 0.115 banks per square mile in less-urban tracts, and 0.004 to 0.14 banks per square mile in many rural tracts. These thresholds can be used to measure and compare bank access in communities across the country.

	Further analysis, using the tract-level measure of Hegerty (2019a), compares socioeconomic deprivation in the bottom five percent of tracts, as measured for adjusted bank density, with the remaining 95 percent. While most deciles’ differences are significant, particularly when radii of five miles are larger are applied, deprivation scores are higher and group differences are largest in the densest tracts. This suggests that more suburban and rural “banking deserts” might not face the same hardships as do their urban counterparts. This finding, which nonetheless is in line with findings mentioned above, requires further investigation.

	These findings point to four additional research directions. First, the impact of bank branch closings can be assessed by incorporating additional years into a study. While FDIC data go back as far as 1994, the quality of their geographic information are limited, and a large percentage of address locations would need to be geocoded by the researcher. Carefully ensuring the comparability of data over time would allow for an analysis of which socioeconomic variables are most closely related to these closings. Second, a finer breakdown of density ranges, whether in terms of more groups, varying thresholds, or a more even distribution of tracts within groups, might also be useful. Third, incorporating geography—especially including place-level information for cities and suburbs—might also enhance the results. Finally, additional socioeconomic variables, including racial makeup, can be included in a multivariate model. The current results, however, provide an interesting look into how exactly one can define a “banking desert.”

\nocite{Hegerty2016}
\nocite{Hegerty2019a}
\nocite{Brennan2011}
\nocite{Degryse2005}
\nocite{Smith2008}
\nocite{Nguyen2019}
\nocite{Chen2014}
\nocite{Ergungor2010}
\nocite{Cover2011}
\nocite{Dunham2018}
\nocite{Hegerty2019b}
\nocite{Kashian2018}
\nocite{Kubrin2011}
\nocite{Woodstock2019}
\nocite{Fed2019}
\nocite{Eisenberg2018}
\nocite{Wheatley2010}

\bibliographystyle{abbrv}
\bibliography{BDpaper}

\end{document}